\documentclass[twocolumn,prl,aps,showpacs,floatfix,superscriptaddress,preprintnumbers,amsmath,amssymb]{revtex4}

\usepackage{graphicx}
\usepackage{endnotes}
\usepackage{pdfpages} 

\begin{document}

\newcommand{\hto}{Ho$_2$Ti$_{2}$O$_7$}
\newcommand{\dto}{Dy$_2$Ti$_{2}$O$_7$}
\newcommand{\ooo}{$[111]$}
\newcommand{\gooo}{$\langle111\rangle$}
\newcommand{\dmdb}{${\rm d}M/{\rm d}B$}

\title{First order metamagnetic transition in {\hto} observed by vibrating coil magnetometry at milli-Kelvin temperatures}

\author{C. Krey}
\affiliation{Technische Universit\"at M\"unchen, Physik-Department E21, D-85748 Garching, Germany}

\author{S. Legl}
\affiliation{Technische Universit\"at M\"unchen, Physik-Department E21, D-85748 Garching, Germany}

\author{S. R. Dunsiger}
\affiliation{Technische Universit\"at M\"unchen, Physik-Department E21, D-85748 Garching, Germany}

\author{M. Meven}
\affiliation{Technische Universit\"at M\"unchen, Forschungsneutronenquelle FRM II, D-85748 Garching, Germany}

\author{J. S. Gardner}
\affiliation{Department of Physics, Indiana University, Bloomington, IN 47408, USA}

\author{J. M. Roper}
\affiliation{Los Alamos National Laboratory, Los Alamos, USA}

\author{C. Pfleiderer}
\affiliation{Technische Universit\"at M\"unchen, Physik-Department E21, D-85748 Garching, Germany}

\date{\today}

\begin{abstract}
We report vibrating coil magnetometry of the spin ice system {\hto} down to $\sim0.04\,{\rm K}$ for magnetic fields up to 5\,T applied parallel to the {\ooo} axis. History dependent behavior emerges below $T_0^*\sim 0.6\,{\rm K}$ near zero magnetic field, in common with other spin ice compounds. In large magnetic fields we observe a magnetization plateau followed by a hysteretic metamagnetic transition. The temperature dependence of the coercive fields as well as the susceptibility calculated from the magnetization identify the metamagnetic transition as a line of first order transitions terminating in a critical endpoint at $T^*_m\simeq 0.37\,{\rm K}$, $B_{m}\simeq1.5\,{\rm T}$. The metamagnetic transition in {\hto} is strongly reminiscent of that observed in {\dto}, suggestive of a general feature of the spin ices.
\end{abstract}

\pacs{75.30.Kz; 75.60.Ej; 75.40.Cx; 75.40.Gb}

\vskip2pc

\maketitle
Metamagnetism (MMT), where a paramagnet undergoes a first order phase transition to a ferromagnetic (FM) state in high 
magnetic fields with a discontinuous jump in the magnetisation, is a pervasive phenomenon in systems based on rare earth or 
transition metals. However, despite striking similarities in the magnetization of a wide range of MMTs their microscopic 
origin may be radically different (see e.g. \cite{Haen87,Thess97,Uhlarz04,Perry01,Iijima90,goto01}). A prominent example of 
such a MMT has been reported recently for {\dto}, where the transition is preceded by a plateau in the magnetization and a 
critical end-point is located at $T_m\sim0.36\,{\rm K}$ and $B_m\sim0.9\,{\rm T}$ for fields strictly parallel to 
a {\gooo} axis \cite{Saka03,Aoki04}. It has been argued that the MMT in {\dto} reflects directly the nature of the 
spin excitations from the zero-field spin state \cite{Cast08}.  The magnetic ions in {\dto} reside on the vertices of a 
pyrochlore lattice of corner sharing tetrahedra in the presence of strong local {\gooo} crystalline anisotropy and 
effective ferromagnetic interactions, leading to geometric frustration.  The ground state is characterized by a residual 
entropy quantitatively comparable to the value of water ice \cite{Rami99,Bram02,Ging09}. This reflects spin disorder at 
low temperature such that two spins are constrained to point outward and two spins in towards the center of a 
tetrahedron \cite{Harr97,Bram02}. 

The MMT in {\dto} has been explained in terms of an entropy reduction which takes place in two steps~\cite{Isak04}. 
First, the system partially magnetises retaining the two-in, two-out state, in which one of the four spins on each 
tetrahedron has a component of the moment antiparallel to the field. Second, as the field increases further, the 
nearest-neighbor spin ice model predicts an ice-rule breaking spin flip to the ‘‘three-in, one-out’’ 
(‘‘one-in, three out’’) state~\cite{Fuka02,Matsu02}.  Recent theoretical work suggests that the spin-flips related to 
the second step may be viewed as emergent magnetic monopoles that condense at the MMT \cite{Ryzh05,Cast08}. This 
scenario was subsequently found to be consistent with the entropy reduction inferred from the magnetocaloric 
effect \cite{Hiro03,Aoki04}, as well as the evolution of the spin relaxation time in the ac 
susceptibility~\cite{Snyd03,Jaub09}, at least above 1\,K. The strong temperature dependence of the specific 
heat below 1\,K \cite{Morr09}, the heat transport \cite{Klem11} and finally magnetization avalanches in low 
magnetic fields \cite{Slob10} are also thought to be consistent with magnetic monopoles. On a microscopic level, 
evidence for magnetic monopoles in {\dto} has been inferred from the observation of lines of reversed spins 
between monopole pairs (cf. Dirac strings) using neutron scattering \cite{Morr09}.

A second candidate for emergent magnetic monopoles is {\hto}. The experimental situation in {\hto} is, however, 
much less clear. Neutron scattering at low temperatures reveals pinch-points in the structure factor consistent 
with power law correlation functions \cite{Fenn09,Henl09}. However, neutron spin echo and neutron backscattering 
experiments \cite{Ehle03,Ehle04,Clan09} suggest intrinsic spin relaxation times much faster than for {\dto}, 
raising the question how this may be reconciled in terms of magnetic monopoles. Further, 
the specific heat displays large nuclear hyperfine contributions~\cite{Bram01}, which manifest themselves as a 
Schottky anomaly, complicating a comparison with the predictions for spin ice behaviour.  
The magnetization of {\hto} for a field along {\ooo} reported so far down 
to 0.5\,K \cite{Petr03,Petr11} showed a non-linear increase around 2\,T reminiscent of {\dto} (distinct from 
discussions of a liquid-gas transition for fields along $[100]$ \cite{Harr98}).  In general,  
the magnetic phase diagram of Ho-based compounds may show strong effects below 0.5\,K due to hyperfine 
interactions as e.g., for the transverse field Ising magnet LiHoF$_4$~\cite{Schecht}.  Hence, detailed measurements well 
below 0.5\,K are needed to explore the putative equivalence between {\hto} and {\dto} quantitatively. 

The observation of history dependencies and dynamics on long timescales in {\dto} and {\hto} in previous studies 
imposes strong experimental constraints. First, tiny sample movements in the applied magnetic field must be 
avoided; they might change the magnetic state. Second, the sample must be rigidly anchored thermally, since 
changes of the magnetization may cause large associated entropy release and uncontrolled local heating effects. 
Third, as the magnetic properties are sensitive to the precise field value and orientation~\cite{Sato06,Sato07} a 
uniform applied field is essential. 

All of these requirements are met by the vibrating coil magnetometer (VCM)~\cite{Legl10,Legl10a} we used to 
measure the magnetization of {\hto}. Data reported in the following between 0.04\,K and $\sim1.8\,{\rm K}$ correspond 
to the properties for field parallel to {\ooo} within a few tenths of a degree. For larger 
misalignments $(\sim \pm 2 ^{\circ })$ we found that the features of interest tend to broaden and shift, 
with additional hysteretic features suggesting complexities beyond the scope of our study. All temperature dependent 
data were recorded while continuously heating at a rate of 5\,mK/min, where zero-field cooled properties and field-cooled 
properties are distinguished. Likewise, all measurements as a function of magnetic field followed one of two well defined 
protocols, denoted as (A) and (B). Details of the temperature dependent measurements as well as the two 
protocols are presented in the supplementary material.

The {\hto} single crystal studied was grown by means of optical float zoning at LANL in an argon 
tmosphere at 5\,mm/hr.  Single crystal neutron diffraction at HEIDI (FRM II) establish a room temperature lattice 
constant of a=10.13(2) \AA\ consistent with previous studies~\cite{Subr83}. A psi scan of the $(555)$ reflection 
confirms the homogeneity and high quality of the sample. Demagnetising fields were accounted for by approximating the 
disc shaped single crystal as an ellipsoid $(7.3 \times 4.8 \times 1.2$ mm$^3$) with a demagnetising 
factor N=0.75 ({\ooo} direction perpendicular to the plane) \cite{demag1,quilliam}.  The edges of the sample were 
wedge shaped, which may result in a distribution of internal fields over a small volume fraction.  This does not 
affect the main conclusions reported here. 

\begin{figure}
\includegraphics[width=0.45\textwidth]{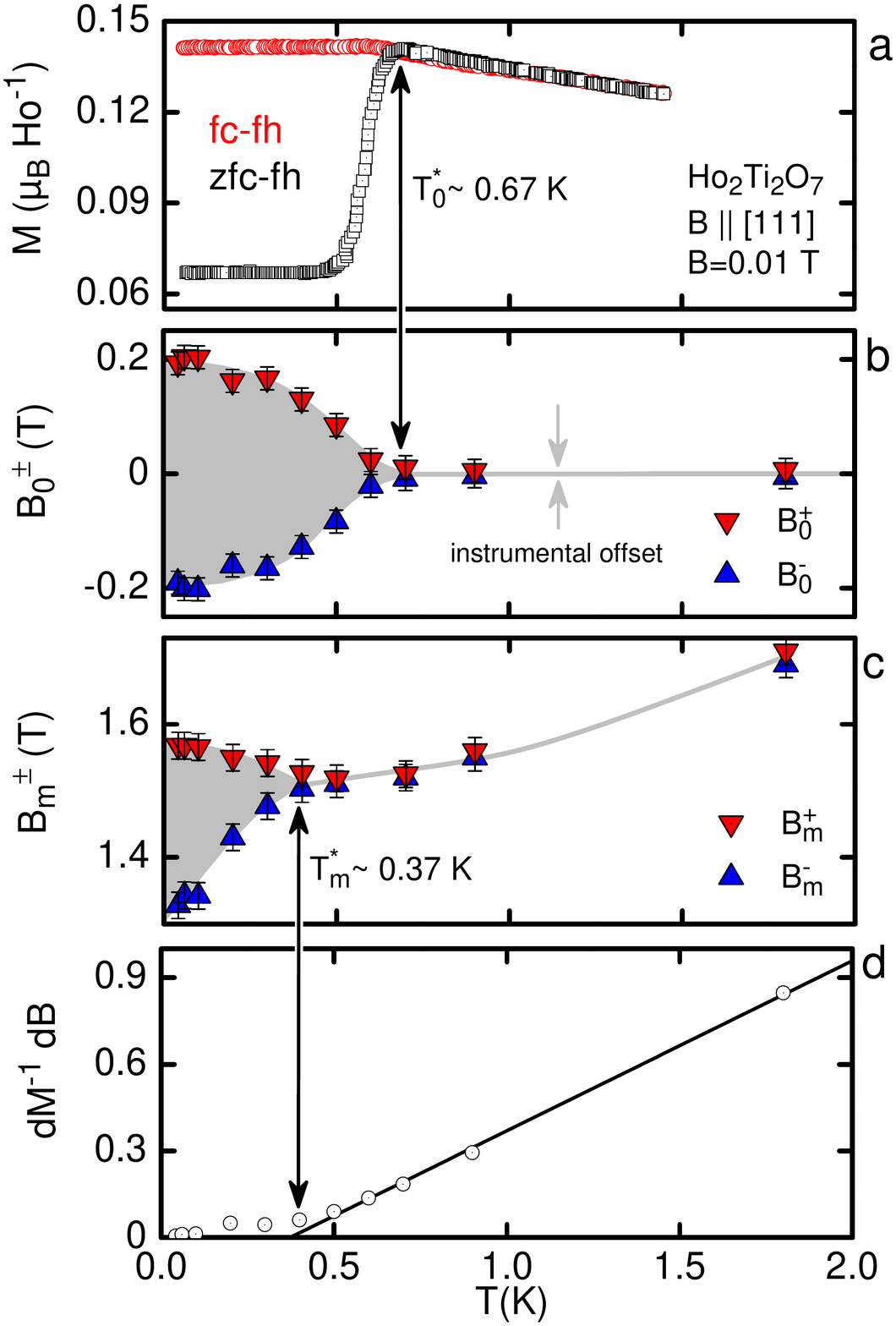}
\caption{(Color online) Characteristic features of the magnetization of spin freezing, hysteresis and a metamagnetic 
transition ending in a critical point in {\hto}. 
(a) Temperature dependence of the magnetization of {\hto} for {\ooo} in a small applied field of 0.01\,T. 
(b) Temperature dependence of the coercive fields with respect to zero field.
(c) Temperature dependence of the coercive fields at the metamagnetic transition. 
(d) Temperature dependence of the peak value of the inverse susceptibility with decreasing temperature, approaching the metamagnetic transition.\\
}
\label{Fig1}
\end{figure}

Shown in Fig.\,\ref{Fig1}\,(a) is the temperature dependence of the magnetization of {\hto} for  $B=0.01\,{\rm T}$ 
along {\ooo}. With decreasing temperature the low-field magnetization increases gradually, characteristic of a 
paramagnetic state. Below a temperature $T_0^*\sim0.6\,{\rm K}$ the zfc/fh and fc/fh data begin to show pronounced 
differences. While the former rapidly decreases to a very low value, the latter remain constant. This behaviour is 
strongly reminiscent of other spin ices, where $T_0^*$ is roughly 0.6-0.75 K for A$_2$B$_2$O$_7$ 
(A=Dy,Ho; B=Ti, Sn)~\cite{Mats00, Snyd04a,Petr11}, i.e., $T_0^*$ is {\it not} proportional to the rare earth 
exchange coupling $J$~\cite{Hert00}.  The microscopic origin of the history dependence shares many features of 
magnetic blocking, but is not understood. For instance, Monte Carlo simulations predict a 1st order transition at 
0.18K \cite{Melk01} which has never been observed experimentally. Interestingly, however, an exponential slowing down 
of the spin relaxation has been reported in solid paramagnets which proceeds via energy levels caused by crystalline or 
hyperfine splitting of the ground state of the ion~\cite{Atsa88}. 

\begin{figure}
\includegraphics[width=0.45\textwidth]{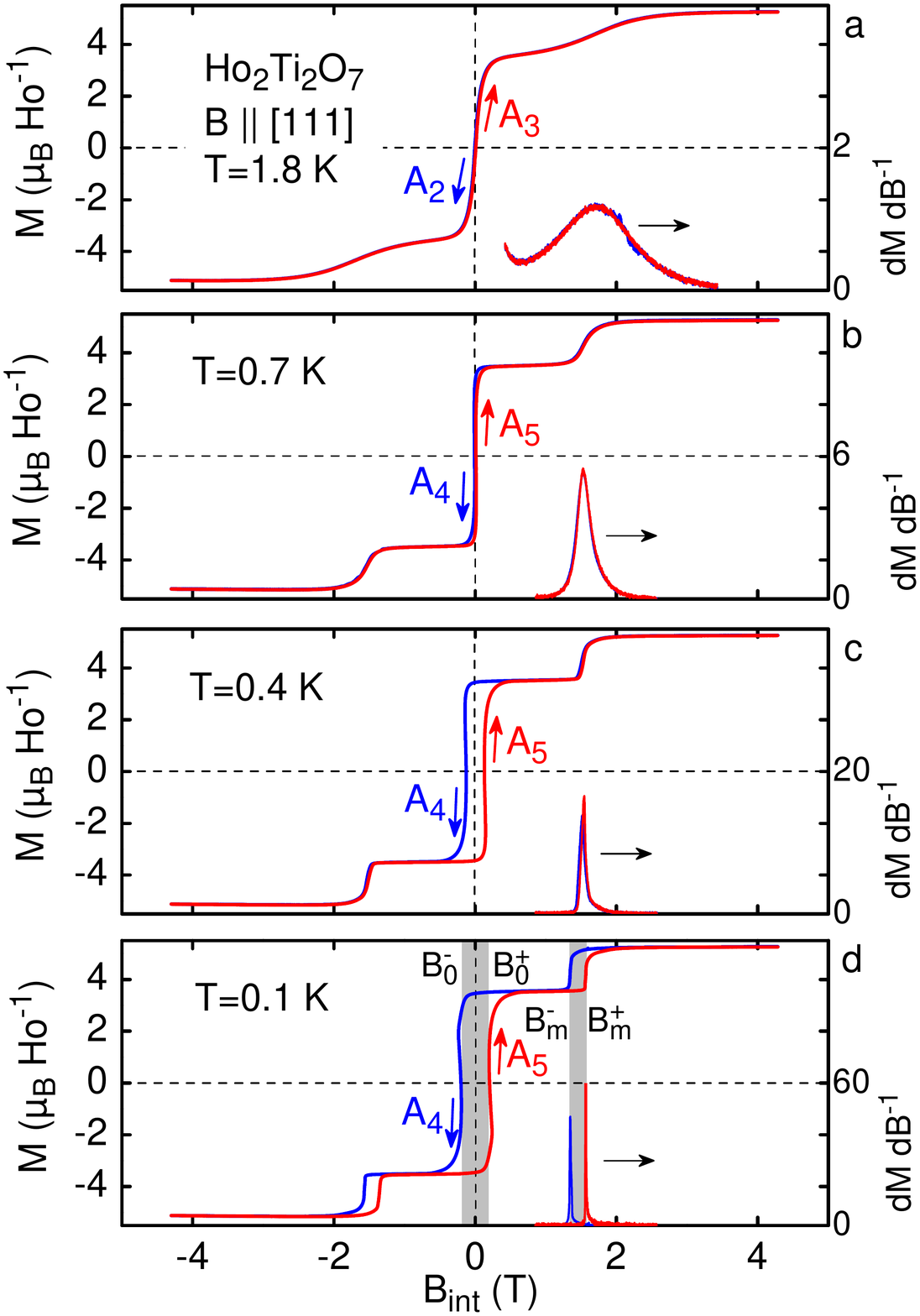}
\caption{(Color online) Magnetic field dependence of the low temperature magnetization of {\hto}. With decreasing 
temperature, hysteresis emerges at $B=0$ and around 1.5\,T. The susceptibility calculated from the magnetization is 
shown on the right hand side of each panel. At the metamagnetic transition it increases strongly with decreasing 
temperature.}
\label{Fig2}
\end{figure}

As a function of magnetic field the history dependence below $T_0^*$ is connected with strong hysteresis with 
respect to $B=0$, followed by a metamagnetic increase at a field $B_m^{\pm}\sim1.5\,{\rm T}$, which becomes distinctly 
hysteretic at low temperatures. This is illustrated in Fig.\,\ref{Fig2}, where typical data are shown as a 
function of \textit{internal} magnetic field, $B_{\rm int}$, given by $\mu_0 H+M$. Data at 1.8\,K shown 
in Fig.\,\ref{Fig2}\,(a) are in excellent agreement with Refs.\,\cite{Petr03,Petr11}. The variation of the magnetization 
near $B=0$ and around 1.5\,T becomes much steeper below $T_0^*=0.7\,{\rm K}$ as shown in Fig.\,\ref{Fig2}\,(b).  
There are several regimes: essentially no hysteresis is observed at 0.7\,K within the accuracy of our set-up. 
At $T=0.4\,{\rm K}<T_0^*$, shown in Fig.\,\ref{Fig2}\,(c), sizeable hysteresis may be seen with respect to zero field, 
while the magnetization at the second step rises non-hysteretically, but more steeply than at 0.7\,K. Finally, as 
shown in Fig.\,\ref{Fig2}\,(d), hysteresis exists with respect to both $B=0$ and $B_m=1.5\,{\rm T}$ for 0.1\,K. 

In order to track the width of the hysteresis loop we define coercive fields $B^-_{0}$, $B^+_{0}$ 
and $B^-_{m}$, $B^+_{m}$ (see Fig.\,\ref{Fig2}\,(d)). Shown in Fig.\,\ref{Fig1}\,(b) are the coercive 
ields $B^-_{0}$, $B^+_{0}$, which increase strongly below $T_0^*$ with decreasing temperature and appear to level 
off around $\pm0.2\,{\rm T}$ as $T\to0$. In contrast, the hysteretic behaviour at high fields appears 
at $T_m^*=0.37\,{\rm K}$, well below $T_0^*$ (Fig.\,\ref{Fig1}(c)). 

\begin{figure}
\includegraphics[width=0.4\textwidth]{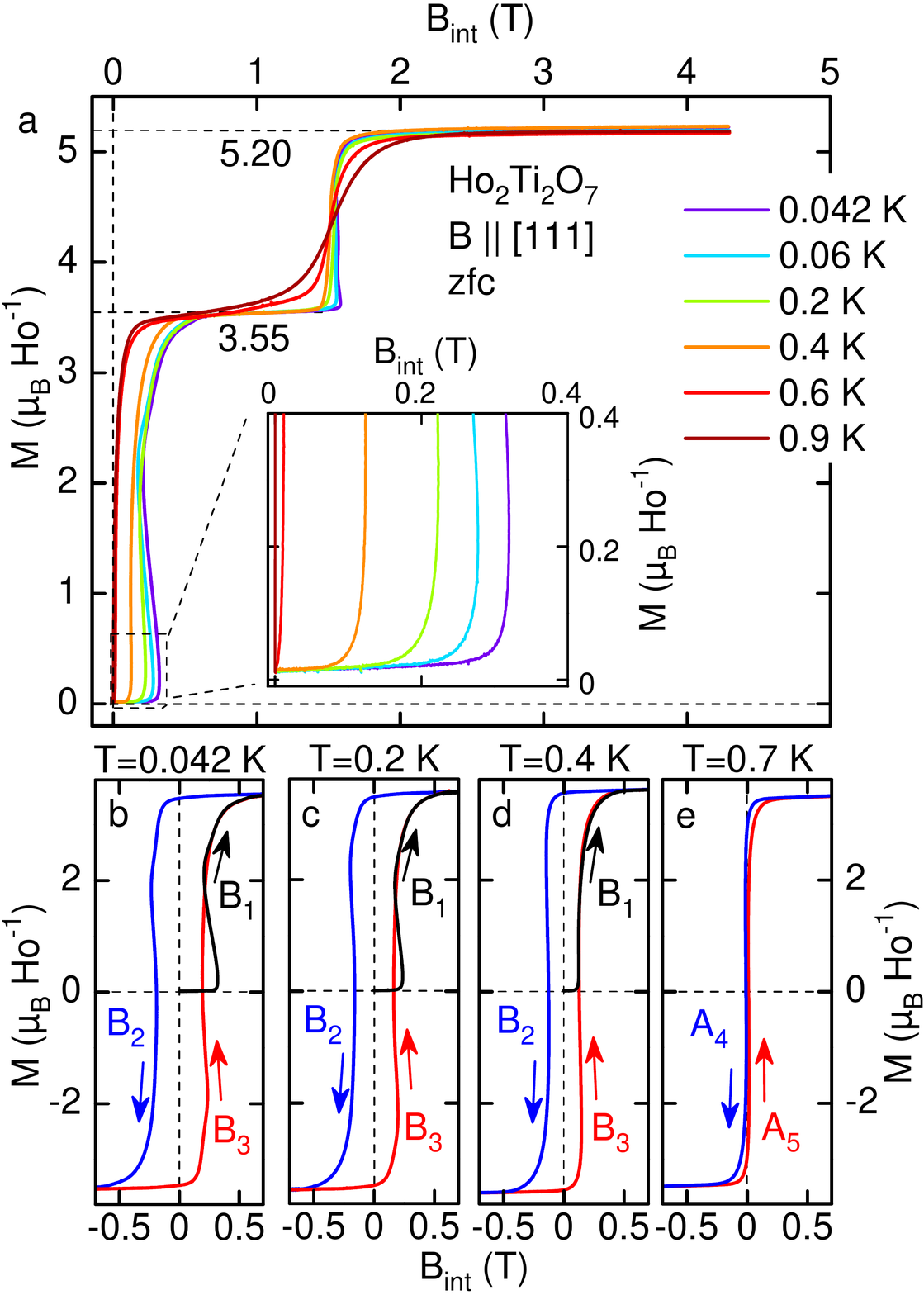}
\caption{(Color online) magnetization of {\hto} as a function of internal magnetic field after correction of the 
demagnetising fields, following field sweeps of protocol (B). For $T<T_0^*$ the initial change of the magnetization is 
zero within experimental sensitivity and followed by a very pronounced increase with increasing field. (a) Magnetic 
field dependence in sweeps of type B1. (b) through (d) magnetization in field cycles of sweep types B1, B2 and B3 (e) 
magnetization in field cycles of sweep types A4 and A5.\\
}
\label{Fig3}
\end{figure}

No conventional long range thermodynamic ordering transition is reported as a function of temperature
at $B=0$, despite the appearance of strong hysteresis. Specifically, there is no development of magnetic Bragg peaks.  
Rather, the notion of strong magnetic blocking is beautifully 
illustrated by the initial field dependence of the magnetization for $T<T_0^*$ after zfc. Shown in Fig.\,\ref{Fig3}\,(a) 
is the magnetization in the first field sweep after zero-field cooling (referred to as B1). Up to field values exceeding 
the coercive fields seen in field sweeps (B2), (B3) the magnetization is unchanged and trapped in the zfc state. The 
peculiar field dependence, specifically the negative slope of the magnetization at intermediate fields is 
reminiscent of {\dto} (Fig.\,\ref{Fig3}\,(b) through (e)), where it has been interpreted as evidence of monopole 
avalanches \cite{Slob10}. However, similar features have also been reported in mesoscopic spin systems~\cite{meso}. 
In view of the enormous sensitivity to the precise orientation of the sample and the large demagnetising factor, as 
well as a small distribution of internal fields at its fringes, a detailed account of possible magnetization avalanches 
is beyond the scope of our study.

Two arguments establish that the hysteresis at $B_m$ is connected with a thermodynamic phase transition, in contrast 
to the behaviour in zero field. First, the inverse susceptibility at $B_m$ calculated from the magnetization essentially 
displays a Curie dependence and vanishes within experimental accuracy at $T_m^*=0.37 (0.01)\,{\rm K}$ 
and $B_m^*=1.52(0.01)\,{\rm T}$ as shown in Fig.\,\ref{Fig1}\,(d). This is characteristic of a critical point 
at $(T_m^*, B_m^*)$. Moreover, the hysteresis observed for $T<T_m^*$ provides evidence that this critical point is 
located at the end of a line of first order transitions.   Interestingly, $T_m^*$ is found to be very close to the 
value for {\dto}  \cite{Saka03}. By contrast, theory predicts scaling of $B_m^*$ with the effective exchange 
coupling $J_{\rm eff}$ \cite{Moes03, Hiro03}.  Given $J_{\rm eff}=5D/3 + J/3$, where $D$ is the dipolar coupling 
between the rare earth moments, and using $J_{\rm eff}/k_{\rm B}=1.1\,{\rm K}$ for {\dto} and 1.83\,K 
for {\hto} \cite{Bram02}, the phase boundary for the latter is predicted to occur at 1.49\,T.  The agreement 
between theory and experiment is very satisfying.

\begin{figure}
\includegraphics[width=0.4\textwidth]{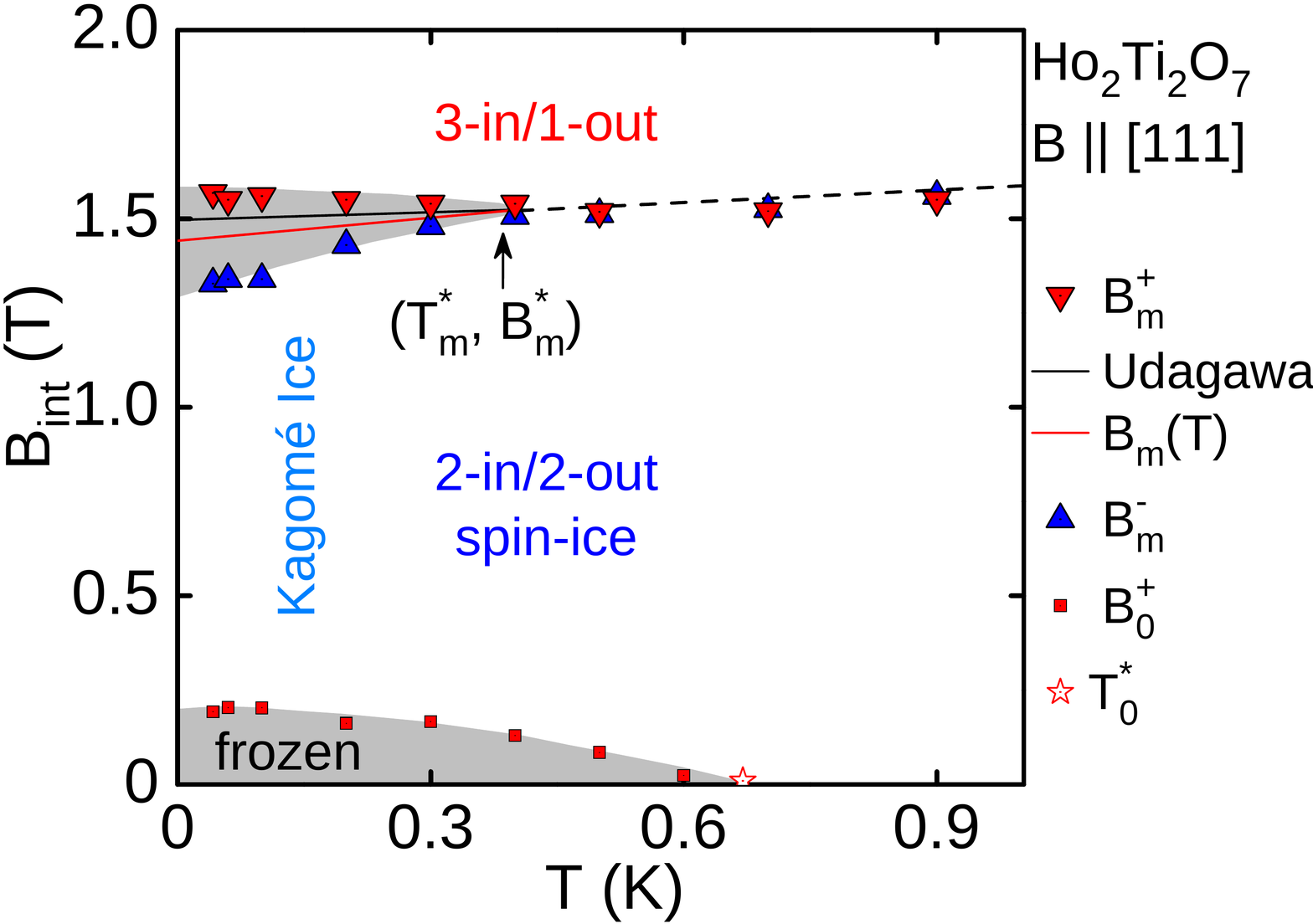}
\caption{(Color online) Magnetic phase diagram of {\hto} for magnetic field parallel to {\ooo}. With 
respect to $B=0$ strong freezing and hysteresis emerges in the magnetization below $T_0^*=0.67\,{\rm K}$ in the 
presence of the residual dynamics seen microscopically \cite{Ehle03,Ehle04,Clan09}. At high fields a line of first 
order metamagnetic transitions terminates in a critical endpoint. The metamagnetic transition separates Kagom\'{e} ice 
behaviour from a three-in/one-out configuration when going from below to above $B_m$.\\
}
\label{Fig4}
\end{figure}

Second, following the Clausius-Clapeyron equation (CC), ${\rm d}B_m/{\rm d}T  = -{\Delta S}/{\Delta M}$, the 
transition at $B_m$ is connected with a strong entropy reduction of a thermodynamic phase transition subject to the 
value of ${\rm d} B_m/{\rm d} T$. Because the transition at $B_m$ is hysteretic determination of the phase boundary 
using CC depends on the choice of ${\rm d} B_m/{\rm d} T$. As a first step it is instructive to suppose Kagom\'{e} ice 
behaviour for $B<B_m$. In this case a rigorous calculation predicts an entropy 
reduction of $0.672^{-1}\,{\rm mol}^{-1}$-ion \cite{Udag02}. The corresponding value 
of ${\rm d} B_m/{\rm d} T=0.073\,{\rm T/K}$, as shown by the solid black line passing through $(T_m^*, B_m^*)$ in 
Fig.\,\ref{Fig4} is perfectly consistent with our data and follows the cross-over line for $T>T_m^*$. However, the 
phase boundary would lie asymmetrically with respect to $B_m^+$ and $B_m^-$.  If we assume instead that $B_m$ is 
located midway in the hysteretic field range or at the field of the peak of {\dmdb} we 
find ${\rm d} B_m/{\rm d} T\approx 0.2\,{\rm T/K}$. In this case the expected entropy release at $B_m$ exceeds the 
entropy associated with Kagom\'{e} ice by an unphysically large factor of 3.65.  Hence the latter prescription must be 
inappropriate.

Strong support for the assumption of Kagom\'{e} ice behaviour for $B<B_m$ may be seen in the magnetization, which 
forms a very well defined plateau at $3.55\,\mu_{\rm B}\,{\rm Ho^{-1}}$ (cf Fig.\,\ref{Fig3}\,(a)). This is quantitatively 
in excellent agreement with theoretical prediction. Moreover, above $B_m$ the magnetization assumes a well-defined value 
of $5.20\,\mu_{\rm B}\,{\rm Ho^{-1}}$ for $T\to0$ characteristic of a three-in/one-out spin configuration. Thus the 
magnetization of {\hto} is quantitatively comparable with that of {\dto} and consistent with the dipolar spin ice model.  
The temperature dependence of $B_m^+$ and $B_m^-$ and the magnetization provide unambiguous evidence of a first order 
transition with a large entropy reduction. Experimental determinations of the residual entropy of the Kagome ice state 
inferred from the specific heat in \dto\ for $B<B_m$ are scattered 
between 0.44(8) -- 0.8(1) J K$^{-1}$ mol$^{-1}$-Dy~\cite{Higa03,Matsu02}, consistent with the value inferred 
from the magnetization, 0.5\,${\rm J\,K^{-1}mol^{-1}}$-Dy \cite{Saka03}. The temperature dependence of the entropy 
release in the magnetocaloric effect in {\dto} \cite{Aoki04} is attributed to magnetic correlations~\cite{Taba06}.

We finally show in Fig.\,\ref{Fig4} the magnetic phase diagram of {\hto} for the {\ooo} axis inferred from our 
magnetization data. In zero field {\hto} enters a spin ice state with strong spin blocking below $T_0^*$. A moderate 
field stabilises a magnetization plateau characteristic of Kagom\'{e} ice. Further increasing the field results in a 
line of first order metamagnetic phase transitions up to a critical endpoint at $T_m^*=0.37\,{\rm K}$ 
and $B^*_m=1.52\,{\rm T}$. This line of transitions separates Kagom\'{e} ice from the spin polarised 3-in/1-out 
configuration.  While the critical field for $T\to0$ scales experimentally with the $J_{\rm eff}$, in agreement with 
theory,  $T_0^*$ and $T_m^*$ appear to be material independent for {\hto} and {\dto}.  The tuning parameter for these 
energy scales remains to be explored.

In conclusion, the remarkable analogy we observe between the properties of {\dto} and {\hto} establishes the field 
induced liquid-gas like transition, interpreted in terms of the condensation of magnetic monopoles, as a 
more generic phenomenon within spin ice systems. 
However, given the importance of dipolar interactions and the resultant power law correlations~\cite{Henl09} as 
an essential prerequisite for a description of the excitations in terms of magnetic monopoles, it also seems clear that 
further Ising like compounds {\textit not} based on Ho or Dy, which have similar classical magnetic moments, must be 
investigated. In this way, the strength of the dipolar coupling could be controllably tuned.
Perhaps most remarkably, 
the phase boundaries appear to be independent of the strength of the  hyperfine interactions, which are much stronger 
in {\hto}.  It is important to note that the dc magnetization is a measure of the {\it zero} frequency response of the system.  
Subleading transverse interactions $J_{\pm}$ which are material dependent and could lead for instance to differing quantum 
tunnelling amplitudes between spin-ice states are predicted to affect the {\it finite} frequency response~\cite{benton12}.
A complete description of the excitations in terms of magnetic monopoles must account for such differences.

We wish to thank P. B\"oni, R. Moessner, R. Ritz, A. Regnat and C. Franz for support and stimulating discussions. Financial support through DFG TRR80 and DFG FOR960 are gratefully acknowledged. CK acknowledges support through the TUM Graduate School.


\clearpage
\section{Supplementary Material for: First order metamagnetic transition in {\hto} observed by vibrating coil magnetometry at milli-Kelvin temperatures}
In this supplement we present a detailed account of the precise temperature and field history used in our measurements and the terminology used to refer to specific temperature and field sweeps in the main text.

\subsection{Temperature dependence}

We distinguish two types of temperature sweeps, namely zero-field cooled/field heated (zfc/fh) and field-cooled/field heated (fc/fh). In both cases data were recorded while heating the sample continuously at a rate of 5\,mK/min. Data denoted as zfc/fh were recorded after initially cooling the superconducting magnet system and cryostat from room temperature, to guarantee the absence of any remanent magnetic fields. To record data denoted as fc/fh, we recooled the sample to base temperature after the zfc/fh measurement was completed (at roughly 1.5\,K), keeping the magnetic field unchanged. Data were then recorded again while continuously heating. 

\subsection{Field dependence: Protocol A}

For field sweeps following protocol (A) the empty VCM was at first demagnetised and the sample cooled in zero magnetic field to $T=1.8\,{\rm K}$. Data were subsequently recorded in three field sweeps as follows: (A1) $0\rightarrow +5\,{\rm T}$, (A2) $+5\,{\rm T}\rightarrow -5\,{\rm T}$, and (A3) $-5\,{\rm T} \rightarrow +5\,{\rm T}$. At the final field of $+5\,{\rm T}$ the temperature was changed. At each subsequent temperature two field sweeps were carried out, namely (A4) $+5 \,{\rm T}\rightarrow -5\,{\rm T}$ and (A5) $-5 \,{\rm T} \rightarrow +5\,{\rm T}$. 

At temperatures above 0.1\,K all data were recorded while sweeping the field continuously at $15\,{\rm mT/min}$. Below 0.1\,K, to avoid heating effects due to eddy currents, data were recorded in a step mode, holding the magnetic field constant while recording the magnetisation.
\newpage

\subsection{Field dependence: Protocol B}

For measurements following protocol (B) the empty VCM was at first demagnetised and the sample subsequently cooled in zero magnetic field from room temperature directly to the temperature of interest.  Data were then recorded in a sequence of three field sweeps: (B1) $0\rightarrow +5\,{\rm T}$, (B2) $+5\,{\rm T}\rightarrow -5\,{\rm T}$, and (B3) $-5\,{\rm T} \rightarrow +5\,{\rm T}$.  For the next set of measurements following protocol (B) the sample was removed from the VCM and the empty VCM again 
demagnetised.

At temperature below 0.1\,K and for all sweeps of type (B1) data were recorded in a step mode. All other data were recorded while sweeping the field continuously at $15\,{\rm mT/min}$. No differences were observed between (A4), (A5) and (B2), (B3), respectively.

\begin{table}[h]
\begin{tabular*}{\columnwidth}{@{\extracolsep{\fill}}ccll}
   & sweep \# & field values & condition\\\hline
  protocol (A) & A1 & $0\rm\,T\rightarrow 5\rm\,T$ & zfc to $1.8\rm\,K$\\
   & A2 & $+5\rm\,T\rightarrow -5\rm\,T$ & \\
   & A3 & $-5\rm\,T\rightarrow +5\rm\,T$ & \\
   & A4 & $+5\rm\,T\rightarrow -5\rm\,T$ & field-cooled \\
   & A5 & $-5\rm\,T\rightarrow +5\rm\,T$ & field-cooled\\
\hline
  protocol (B) & B1 & $0\rm\,T\rightarrow +5\rm\,T$ & zfc from ambient \\
   & B2 & $+5\rm\,T\rightarrow -5\rm\,T$ & \\
   & B3 & $-5\rm\,T\rightarrow +5\rm\,T$ & 
\end{tabular*}
\caption{Summary of protocol (A) and (B) used in measurements as a function of magnetic field. Zero-field-cooled and field-cooled starting conditions are denoted as zfc and fc, respectively.}
\end{table}
\end{document}